# Using Hashtags to Analysis Purpose and Technology Application of Open-Source Project Related to COVID-19

Liang Tian, Chengzhi Zhang*

Department of Information Management, Nanjing University of Science and Technology, Nanjing 210094, China

**Abstract:** COVID-19 has had a profound impact on the lives of all human beings. Emerging technologies have made significant contributions to the fight against the pandemic. An extensive review of the application of technology will help facilitate future research and technology development to provide better solutions for future pandemics. In contrast to the extensive surveys of academic communities that have already been conducted, this study explores the IT community of practice. Using GitHub as the study target, we analyzed the main functionalities of the projects submitted during the pandemic. This study examines trends in projects with different functionalities and the relationship between functionalities and technologies. The study results show an imbalance in the number of projects with varying functionalities in the GitHub community, i.e., applications account for more than half of the projects. In contrast, other data analysis and AI projects account for a smaller share. This differs significantly from the survey of the academic community, where the findings focus more on cutting-edge technologies while projects in the community of practice use more mature technologies. The spontaneous behavior of developers may lack organization and make it challenging to target needs.

---

* Corresponding author, Email: zhangcz@njust.edu.cn

## 1.0 Introduction

Since discovering the COVID-19 outbreak in late 2019, the pandemic has spread rapidly across the globe. Modern technologies have made essential contributions to the pandemic. Artificial intelligence, the Internet of Things, and big data have contributed to epidemic prevention and control. These technologies are being used for disease diagnosis, contact tracing, surveillance, and social power to reduce the number of COVID-19 infections and better treat patients (Jakhar and Kaur 2020).

Reviewing the application of artificial intelligence, big data, and other technologies in pandemics will help facilitate future research and technology development to provide better solutions for dealing with COVID-19 pandemic and future pandemics (He, Zhang, and Li 2021). A survey of the literature through published academic papers (e.g., Karami et al. 2021; Vaishya et al. 2021), patents (e.g., Keestra et al. 2022), and news (Zhao et al. 2021) is a common way to review the application of technologies in the pandemic. Unlike previous studies, the present study aims to review the application of new technologies in the pandemiby investigation communities of practice. Communities of practice have an active role in knowledge sharing and serve as a bridge between knowledge and technology (Adams and Freeman 2000). GitHub is the world's largest open-source project hosting site. It is an essential community of practice for modern technologies such as artificial intelligence, the Internet of Things, and big data. In recent issues of the Mining Software Repositories (MSR) conference, many studies have been analyzed based on GitHub repositories (Pickerill 2020). As can be seen, the analysis of GitHub repositories is highly representative and therefore is the subject of this paper.

This study will discuss the application of technologies in the Github community to fight against the pandemic from two perspectives. On the one hand, is the functionality and type of COVID-19-related projects. On the other hand, it is the association between different technologies and functionalities. Topics[2] are hashtags the project developer adds to classify the repository, including the intended purpose, subject area, community, or programming language. Observing the trend of hashtags may realize the events in the community or people focus on what topics(Chen and Kao 2015). We collected the information on COVID-19-related projects on GitHub. Since some developers did not label hashtags, we trained a multi-label classification model using the projects with hashtags and extracted hashtags for the projects lacking hashtags. We did co-word clustering, word frequency statistics, and association rules mining for the hashtags to achieve our research objectives. To better illustrate the research goals of this paper, Figure 1 shows a project with a higher star rating. The descriptions in "About" and "Readme" show that this project provides COVID-19 case data API. The hashtags in "About" reflect the primary purpose (API) of the project and the technology (Redis) used in the project.

The main contributions of this paper include the following. Analyzing the response of specific communities of practice to the pandemic will help facilitate better solutions for the community of practice in response to the COVID-19 and future pandemics. Summarizing the functionality of projects initiated by the IT community of practice during the pandemic and the technologies used will help create a larger pool of pre-existing technologies to address future crises. An examination of GitHub shows the differences in technology adoption during the pandemic between the communities of practice and the academic community. This can provide helpful insight into the rapid adoption of emerging technologies during the pandemic.

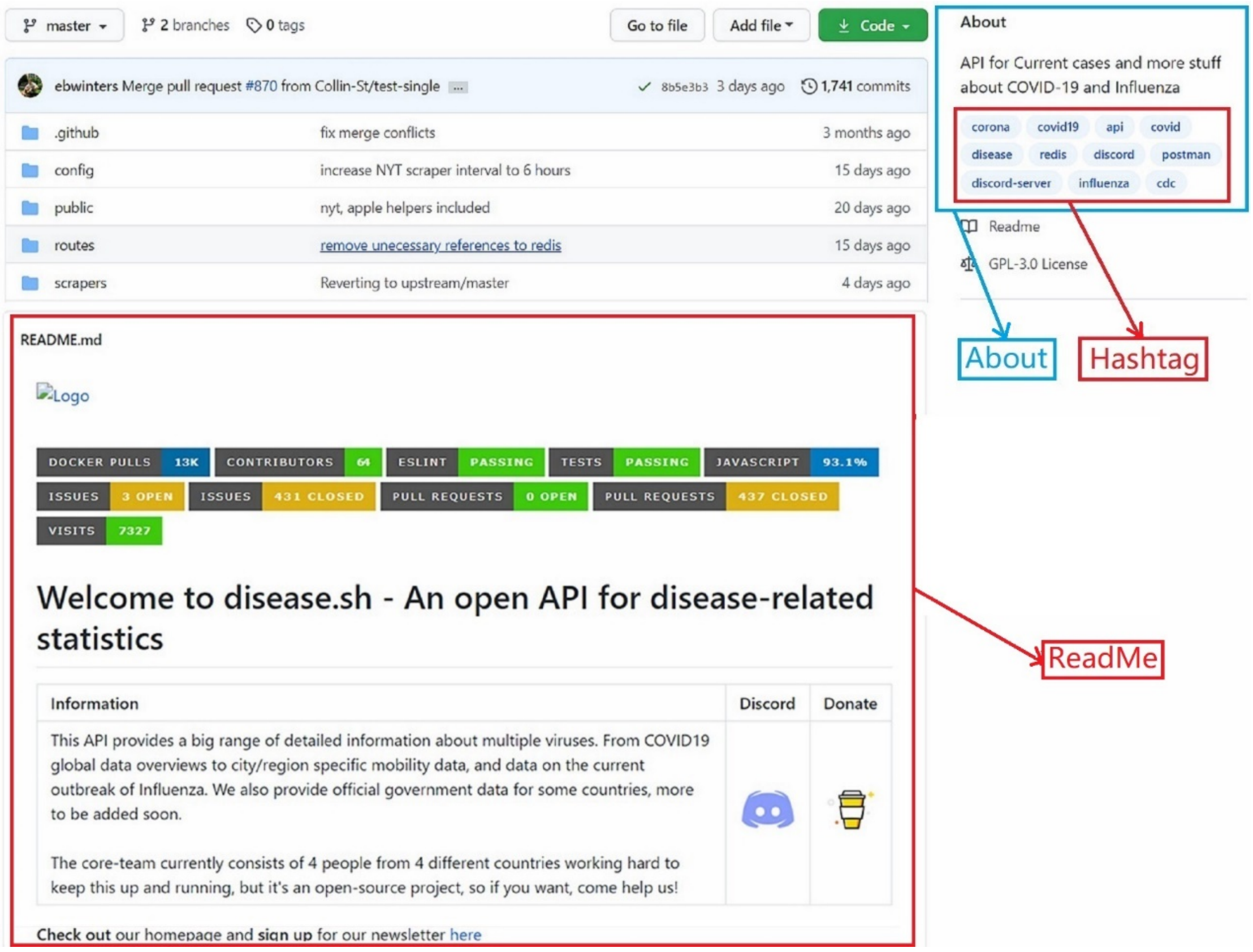

Figure 1. An example of COVID-19-related projects.

## 2.0 Related Works

In this section, we review two aspects of the pandemic. We first check the impact of COVID-19 on society in general and then highlight the importance of technology in the context of a pandemic. We then review relevant studies that have examined the application of technology during the pandemics.

### 2.1 Overview of pandemic impact

The pandemic has profoundly impacted the global economy, industrial production, and lifestyles. Assessing this impact and identifying possible challenges is essential for discovering possible mechanisms to address these issues. Numerous studies have examined the effects of pandemics on different aspects of social life.

By mining and analyzing policy texts and social media data, combined with interviews and other methods, the researchers tried to reveal the pandemic's impacts on social life deeply. Researchers try to analyze the effect of the pandemic on the global economy and seek solutions to the crisis (Song and Zhou 2020). The pandemic can bring terrible disruptions to financial markets, and measuring the predictability of market fears is vital to stabilizing financial markets (Ghosh and Sanyal 2021). The pandemic has forced companies to adjust their strategies and goals (Yadav, Kar, and Kashiramka 2021), creating much uncertainty for employment (Koch, Plattfaut, and Kregel 2021). The pandemic limited travel and made people's work and study quickly shift online, changing the way people live (Chakraborty and Kar 2021).

The impact of COVID-19 on the global economy is undoubtedly devastating. But in this crisis, there are both opportunities and challenges. The pandemic accelerated the adoption of emerging technologies, which contributed to saving lives and improving health during the pandemic (Brem, Viardot, and Nylund 2021). Big data technologies help us predict and control the spread of pandemics (e.g., Srinivasa Rao & Vazquez, 2020). Artificial intelligence technology can achieve rapid patient diagnosis and new drug discovery (e.g., Apostolopoulos & Mpesiana, 2020; Randhawa et al., 2020). Emerging technologies are critical for us to beat the pandemic, and open-source projects guarantee the rapid adoption of these emerging technologies.

### 2.2 Investigation of technology application during the pandemic

Innovations and technology applications have provided new solutions to many challenges during the pandemic. Many researchers have examined technological innovations and applications during the pandemic. Most reviews of technology applications during the pandemic have examined numerous technologies during the pandemic through a survey of academic literature, including Industry 4.0 technologies (Javaid et al. 2020), digital technologies (Wang et al. 2021), artificial intelligence, cloud computing (Alharbi and Abdur Rahman 2021), etc. Data for the survey of the academic literature were primarily obtained from numerous literature databases, including Scopus, Google Scholar, Science Direct, and Research Gate (e.g., Javaid et al. 2020; Vaishya et al. 2021). In addition to the academic literature review, some studies have also reviewed technology applications during the pandemic by examining patents (e.g., Solanki et al. 2021; Alshrari et al. 2022). A few studies have examined the use of technology during the pandemic by reviewing news (e.g., Zhao et al. 2021).

Academic literature and patents are the most direct results of scientific research and reflect the innovation and application of technologies during the pandemic. But there is also a need to transfer knowledge from scientific and technological achievements to production applications, which requires sharing technology, skills, and knowledge from one institution to another (de Wit-de Vries et al. 2019). As an essential tool for knowledge sharing and organizational learning (Marsick, Shiotani, and Gephart 2014), communities of practice serve as a bridge between technology and knowledge. As one of the world's largest IT communities of practice, GitHub launched many open-source projects during the pandemic. These open-source projects share code and allow community members to create and build together. Open strategies can yield higher returns (Barge-Gil 2013), while modern software implementations rely entirely on open-source libraries and components (Shrestha et al. 2020). Community plays a unique role in creating new things (Soos and Leazer 2020); it can be seen that the open-source community is the leading practice frontier of open innovation. Open-source projects provided many open-source libraries and components for the rapid adoption of technologies during the pandemic. Learn how the pandemic "mobilized" specific communities of practice by reviewing the functionality of projects started by the open-source community during the pandemic and the technologies used. This will help the community of practice provide better solutions in response to the COVID-19 pandemic and future pandemics.

## 3.0 Methodology

To discover the functionality of pandemic-related projects in the Github community and apply different technologies, this study uses the Readme and Topics of open-source projects as the data source for analysis. This study first explores the main types of projects through cluster analysis, explores the association between functionality and related technologies through correlation analysis, and analyzes the evolution of open-source projects in the GitHub community during the pandemic. The specific methods of this study mainly include five steps: data collection, data cleaning and preprocessing, synonym hashtags preprocessing, hashtags extraction, and data analysis. The research method is shown in Figure 2.

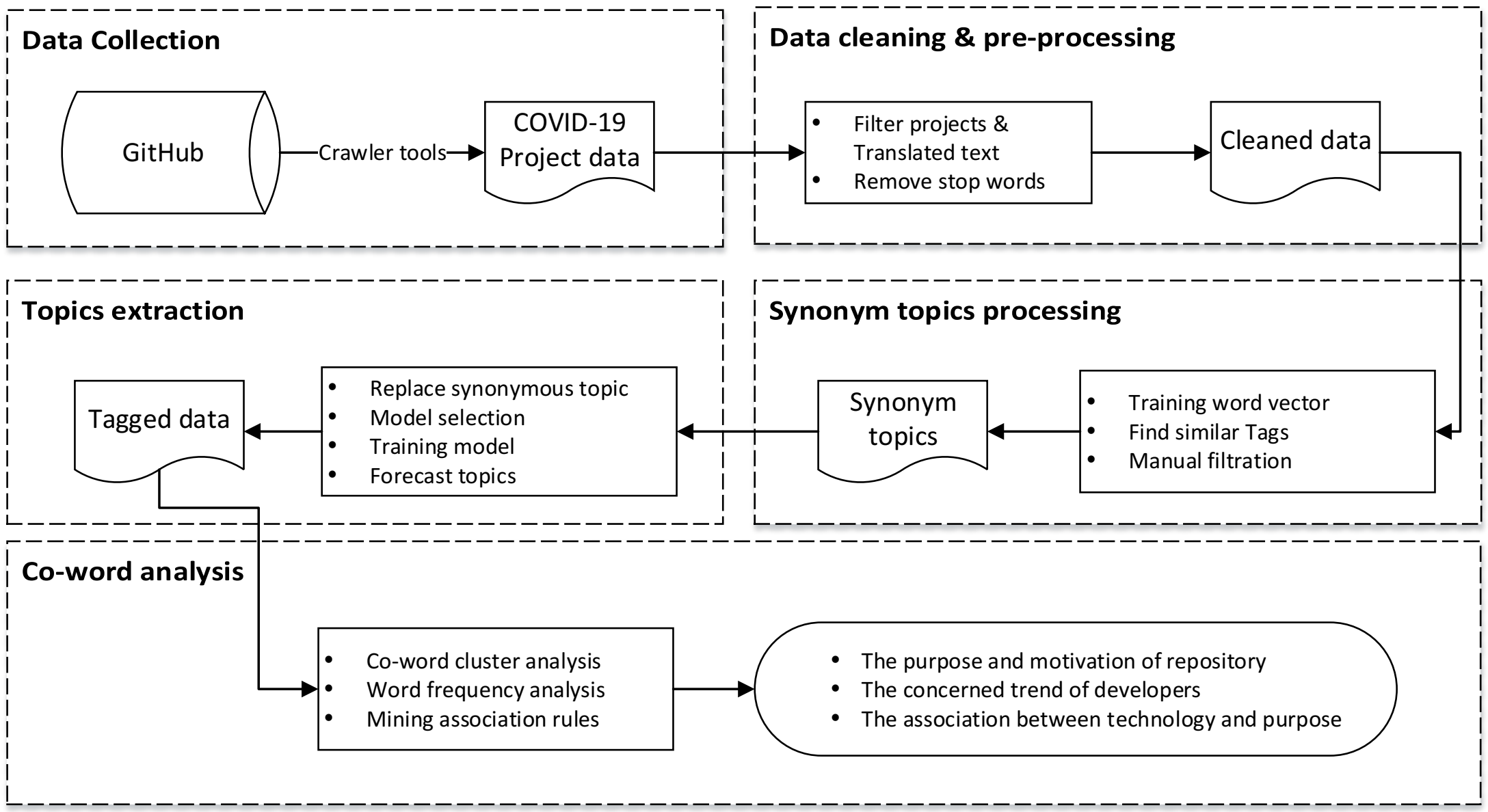

Figure 2. The framework of this study.

### 3.1 Data Collection

GitHub has a large number of low-quality projects that are not managed. These projects cannot be analyzed in this study. The community's response to the project is an appropriate standard to measure the quality of the project. Therefore, this study uses the number of project stars to measure the project quality and limits the number of project stars to be greater than 0. Since the name in the early stage of the epidemic is not clear, we used "2019-ncov", "coronavirus," and "COVID-19" as the search terms to ensure good recall. Limit the project creation time to August 31, 2020, and retrieve it with GitHub's search API[3]. Use the crawler tool to collect the retrieved item data. The data collection time is September 23, 2020, and we ordered 16,706 data.

### 3.2 Data cleaning and pre-processing

This section mainly introduces data cleaning and pre-processing. The functionality of the project and the technology used are hidden in the title, Readme, and hashtag. The title and Readme text were merged into a project introduction to facilitate analysis. Due to the lack of project introduction in some projects, it is difficult for other developers to understand the specific purpose and functionality of the project. This kind of project was not taken as our analysis object, so we screened out 15541 projects whose title and Readme are not empty and whose stars are more significant than 0. The developers in GitHub come from worldwide, so the title and Readme of the project also contain multiple languages[4]. We used Baidu translation API5 to translate all non-English project introductions into English to conduct a unified analysis. The translated project introduction was sorted in descending order according to the number of characters, and the distribution chart of project introduction length, as shown in Figure 3, was obtained.

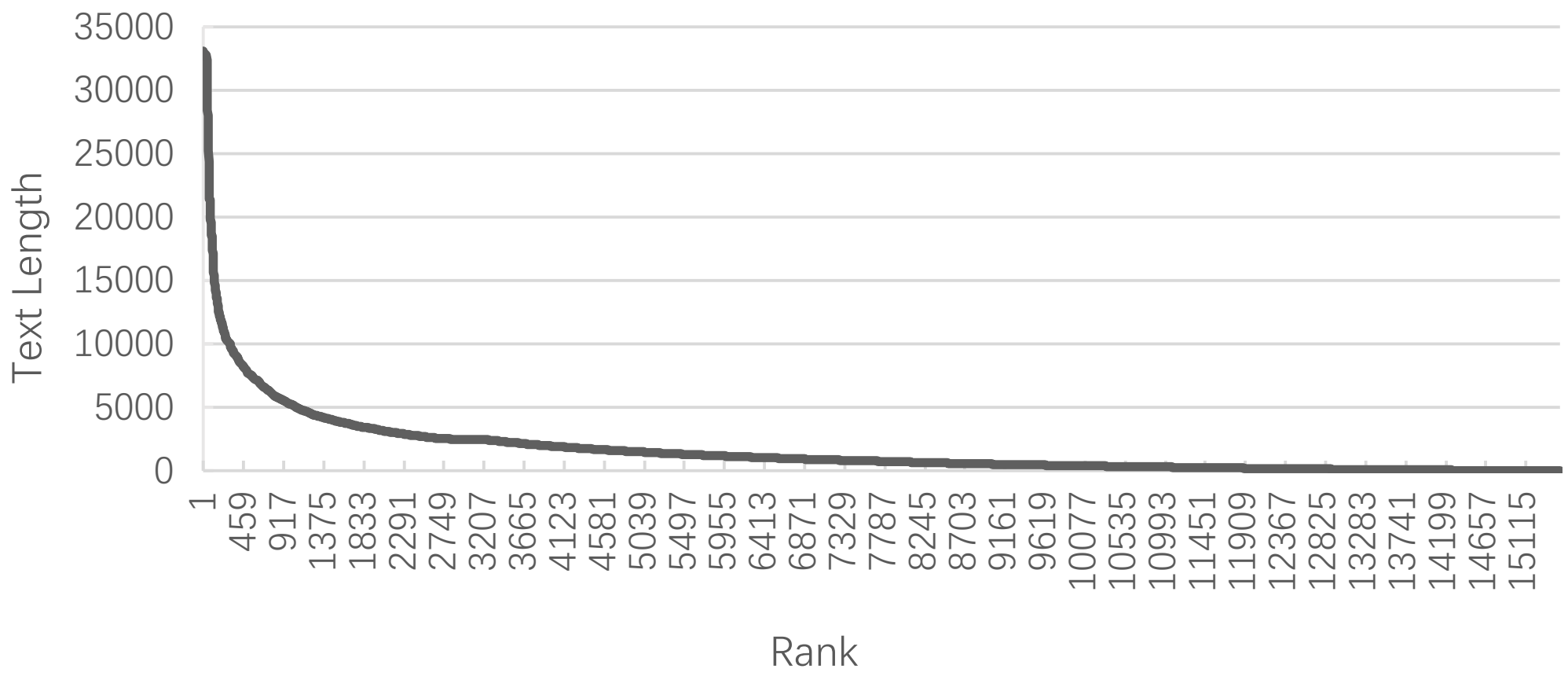

Figure 3. Sorting of the length of the project introduction.

As shown in Figure 3, the length of the project introduction follows the power-law distribution. Because some project introductions are too short to clearly describe the project purpose and other information, these projects need to be removed. Through manual screening, it was found that the introduction was more complete when the introduction length was more significant than 200. Therefore, this study excluded projects with introduction lengths below 200. After screening, a total of 12,199 items were retained. For the remaining tasks, we remove the stop words[6] and other special characters from the introduction (i.e., only English and numbers are kept). The data processing results are shown in Table 1.

| Process | Raw data | Initial screening result | Preprocessing result |
| --- | --- | --- | --- |
| Number of projects without hashtags | 12342 | 11268 | 8359 |
| Number of projects with hashtags | 4364 | 4273 | 3840 |
| Total | 16706 | 15541 | 12199 |

Table 1. Data processing results.

### 3.3 Synonym hashtags processing

Since the project Topics are labeled by different project developers, there is no uniform standard. There will be many synonymous hashtags, which need to replace the synonymous hashtags with a unified hashtag. Synonymous hashtags mainly include the following four situations:

i. Different forms of the same root word, such as “healthy” and “health,” “chatbots,” and “chatbot”.

ii. Spelling errors, such as “anroid” and “android.”

iii. Hyphens exist in the same hashtags, such as "machine learning" and "machine-learning", "Android application," and "Android-application".

iv. Synonymous tags, such as "COVID-19", "novel-coronavirus-2019", and "novel-coronavirus,” express the same meaning.

Given the above four situations, the present study found synonymous hashtags by calculating the semantic similarity and editing distance (Marzal and Vidal 1993) of hashtags and combining them with manual filtering. With the help of the gensim[7], the Word2Vec is used to train the word vectors with a corpus of the introduction of all projects (Mikolov et al. 2013). For each hashtag in the existing topic collection, the ten tags that are most semantically similar to the hashtag and the hashtags less than one-fifth of the word length away from the edit of that hashtag are selected. The similar hashtags automatically screened out are manually screened to find synonymous hashtags and build a synonym dictionary. The number of synonymous hashtags in different situations is shown in Table 2.

| Synonymous hashtag type | Count | Proportion |
|---|---|---|
| Different forms of the same root word | 76 | 16.9% |
| Spelling errors | 43 | 9.6% |
| Hyphens exist in the same hashtags | 145 | 32.2% |
| Synonymous tags | 186 | 41.3% |

Table 2. The number of synonymous hashtags of different types.

### 3.4 Hashtag extraction

One of the main objectives of this study is to analyze the purpose and functionality of the project. We will explore and analyze the intended purpose, the functionality, and the programming language with the help of the project hashtags. In the filtered data of this article, 3840 projects have been tagged by developers, and the remaining 8359 items were not. Therefore, this study will train classification models using labeled data and automatically assign tags to unlabeled data.

There are a large number of low-frequency hashtags. This study selected the core hashtags from the existing collection as candidate hashtags for analysis to reduce the interference of low-frequency words and facilitate visualization. The current hashtags were sorted in descending order of word frequency. As shown in Figure 4, the word frequency ranking of these hashtags presents a power-law distribution. Formula (1) is the calculation formula of cumulative word frequency proportion. The Ci is the word frequency of the ith tag with the highest word frequency. The P is greater than or equal to 50%, and the top k hashtags in descending order of word frequency were selected as the core hashtags.

$$P = \frac{\sum_{i=1}^{k} C_i}{\sum_{i=1}^{n} C_i} \times 100\% \qquad (1)$$

In this paper, we used TfidfVectorizer in sklearn[9] to vectorize the text. This method takes TF-IDF(Salton and Buckley 1988) as feature weight. The present study used chi-square to select features. We tried the commonly used supervised learning models, including SVM(Chang and Lin 2011), NB (Manning et al. 2008), KNN (Keller, Gray, and Givens 1985), Logistic regression (Hosmer et al. 2013), and Random Forest (Breiman 2001). The best model from the above models was selected to predict the data with missing hashtags.

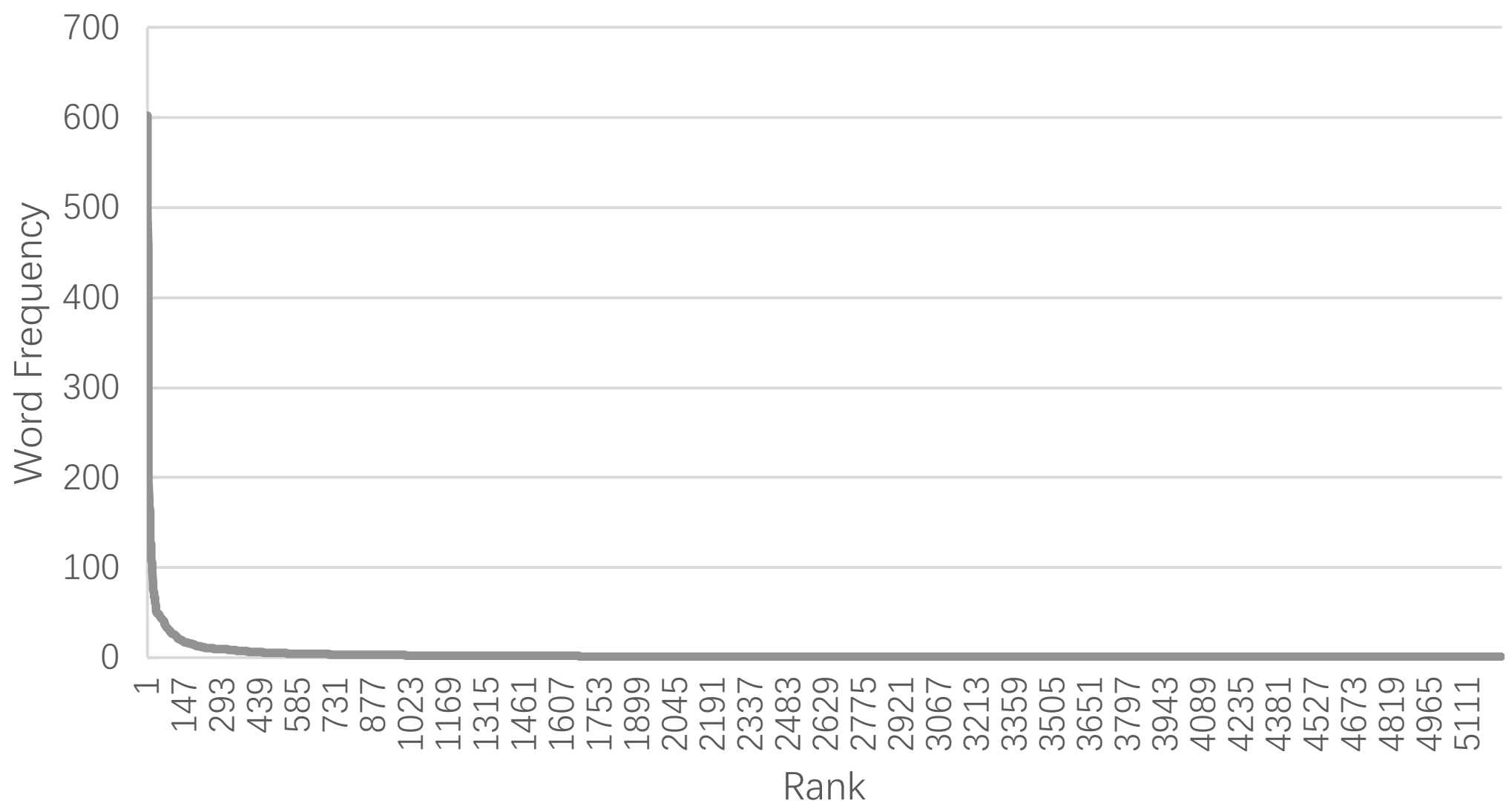

Figure 4. Word frequency sorting of hashtags.

## 3.5 The functionality identification and technology application analytics

This section mainly introduces the method and process of co-word clustering analysis, word frequency statistical analysis, and association rule mining for hashtags to analyze the project purpose, functionality, and related technology application.

3.5.1. Co-word clustering analysis

The project hashtags contain the expected purpose and functionality of the project, and the co-occurrence of different hashtags can reflect the relevance between the project purpose and functionality. Therefore, the co-word cluster analysis was carried out on the project hashtags to understand the primary goals and functionality of the epidemic-related open-source projects. The basic idea of spectral clustering comes from the spectral graph partition theory. Spectral clustering transforms the clustering problem into a graph partitioning problem. The partitioning into subgraphs has the maximum similarity within the subgraphs and the minimum similarity between the subgraphs (Jianbo Shi and Malik 2000). A co-occurrence matrix is created based on the co-occurrence of the labels, and the co-word matrix is converted into an affinity matrix using Ochiia coefficients. The affinity matrix was spectrally clustered using the sklearn[10], and the optimal number of clusters was judged using the Calinski Harabaz (1974) criterion. The more significant the Calinski Harabaz, the better the clustering effect. Input the clustering results into vosViewer[11] for visualization and analyze open-source projects' primary purposes and functionality.

3.5.2. Trend analysis of open-source projects

To explore the functionality of open-source projects in the GitHub community during the pandemic and the problems they aim to solve, and how different types of projects have changed over time. In this paper, we do a descriptive statistical analysis of the average word frequency of hashtags in projects with other purposes. Formula (2) is the calculation formula of intermediate word frequency. $C_i$ is the word frequency of the *ith* hashtag in cluster C, and N is this cluster's total number of hashtags.

$$Ave_Frequency = \frac{\sum_{i=1}^{N} C_i}{N} \quad (2)$$

3.5.3. Association rules mining

To explore the usage of related technologies in different projects, this article used association analysis to mine the association rules hidden between hashtags. Apriori is a widely used association rule mining algorithm proposed by R. Agrawal and r. Srikant(1994). We used Apriori to mine association rules. Based on the specific tasks of this paper, support, confidence, and lift are explained as follows. The support of hashtag set A is the probability that all hashtags in set A co-occur in the hashtag set D. The confidence of association rule A->B is the possibility that hashtag put B is conditional on the occurrence of hashtag set A. The lift of association rule A->B is the confidence of association rule A->B divided by the support of B. The formulas of support, confidence, and lift are (3) (4) (5).

$$Support(A) = \frac{count(A)}{count(D)} = P(A) \quad (3)$$

$$Confidence(A{-}{>}B) = \frac{count(AB)}{count(A)} = \frac{P(AB)}{P(A)} \quad (4)$$

$$lift(A{-}{>}B) = \frac{Confidence(A{-}{>}B)}{Support(B)} = \frac{P(AB)}{P(A)P(B)} \quad (5)$$

A rule with support and confidence greater than the threshold is a strong association rule. Lift is a reflection of whether the association rule is valuable or not. If it is greater than 1, the association rule is useful, and the greater the lift is, the better the association rule is. The association between project functionality and related technologies between technologies and technologies is further analyzed by mining the association in project hashtags.

## 4.0 Results

This section introduces the purpose and functionality of COVID-19 related projects, the evolution trend of open-source projects on GitHub, and the correlation between functionalities and technologies.

### 4.1 Experiment results of hashtags extraction

The task in this article is a multi-label classification task. The multi-label classification task mainly contains two types of evaluation, i.e., label-based and sample-based (Zhang and Zhou 2014). In this paper, the sample-based is chosen as the evaluation metric. The present study used macro precision (P), macro recall (R), and macro $F_1$ to evaluate the model. For a project with missing hashtags, if the number of correctly predicted hashtags is TP, the number of incorrectly predicted hashtags is FP, and the number of unpredicted hashtags is FN. The formulae for the model evaluation metrics are shown in (6), (7), and (8).

$$P = \frac{TP}{TP+FP} \quad (6)$$

$$R = \frac{TP}{TP+FN} \quad (7)$$

$$F_1 = \frac{2 \times P \times R}{P+R} \quad (8)$$

Data with developer-tagged projects were used as the training set, and five-fold cross-validation was performed. The evaluation results of SVM, NB, KNN, Logistic Regression, and Random Forest are shown in Table 3.

| Model | P (%) | R (%) | F1 (%) |
| --- | --- | --- | --- |
| SVM | 72.9 | 54.1 | 62.1 |
| NB | 70.8 | 56.0 | 62.5 |
| KNN | 79.8 | 47.2 | 59.3 |
| LogisticRegression | 46.2 | 63.7 | 53.6 |
| RandomForest | 76.9 | 44.9 | 56.7 |

Table 3. Model performance of hashtags extraction.

It can be seen from Table 3 that NB and SVM can perform better in this task. Since SVM had a higher accuracy rate, we chose SVM to extract hashtags. The top ten projects marked with stars in the classification results are given in Table 4. The URL is the project link, and Hashtags are the classification results. Checking the corresponding items shows that the selected hashtags can more accurately reflect the project's content.

| No. | Title | URL | Hashtags |
| --- | --- | --- | --- |
| 1 | COVID-19 global data (from JHU CSSE for now) as-a-service | https://github.com/mathdroid/COVID-19-api | coronavirus tracking;react;reactjs;covid 19 |
| 2 | Source code of the Beta of the NHS COVID-19 Android app | https://github.com/nhsx/COVID-19-app-Android-BETA | android; covid 19 |
| 3 | The repository contains an ongoing collection of tweets IDs | https://github.com/echen102/COVID-19- | covid 19; Twitter |
| | associated with the novel coronavirus COVID-19 (SARS-CoV-2), which commenced on January 28, 2020. | TweetIDs | |
| 4 | Dados diários mais recentes do coronavírus por município brasileiro | https://github.com/turicas/covid19-br | covid19 data;covid 19 |
| 5 | Using deep learning to generate novel molecules as candidates for binding with coronavirus protease | https://github.com/mattroconnor/deep_learning_coronavirus_cure | machine-learning;covid 19;deep learning |
| 6 | The coronavirus dataset | https://github.com/RamiKrispin/coronavirus | covid19 data;covid 19 |
| 7 | Aspires to help the influx of bioRxiv / medRxiv papers on COVID-19 | https://github.com/karpathy/covid-sanity | covid 19; python |
| 8 | Data from BAG Tweets made useful. | https://github.com/daenuprobst/covid19-cases-switzerland | covid19 data;covid 19 |
| 9 | In standard format | https://github.com/MinCiencia/Datos-COVID19 | covid19 data;covid 19 |
| 10 | Data Science applied to the new coronavirus pandemic. | https://github.com/3778/COVID-19 | epidemiology;covid 19;simulation |

Table 4. Example of extraction results of COVID-19 projects without hashtags on GitHub[12].

### 4.2 The purposes and functionalities of COVID-19 related projects on Github

According to the co-occurrence of hashtags, the results of hashtag extraction were clustered. When the number of clusters was equal to 4, the Calinski and Harabasz score was highest. Therefore, the cluster number was determined as 4. The clustering results were imported into vosViewer for visualization display, and the clustering results are shown in Figure 5. According to the clustering in Figure 5, different colors represent different clusters. It can be seen that the purposes and functionalities of COVID-19 related open-source projects on GitHub mainly include the following

four contents:

i. Application of AI. In the blue part of Figure 5, the functionalities of the projects in this category mainly include the application of AI in the pandemic. Some projects use deep learning, machine learning (machine learning, deep learning), and other technologies to assist the diagnosis of patients with clinical data. Some projects use computer vision technology for mouthpiece detection, control of social distance, etc. In addition, some of the projects provide information sharing portals for drug discovery, the use of machine learning methods to identify potential drugs, etc.

ii. Application of data science. In the green section of Figure 4, the functionalities of this category are focused on the various stages of the data science lifecycle. It includes data collection, storage, modeling, analysis, and data visualization. Data collection mainly consists of some web page-grabbing tools. Data storage related projects include open repositories directly providing relevant data, and some projects provide data API. Data modeling related projects mainly focus on pandemic development data combined with infectious disease models to predict pandemic trends. Data analysis and data visualization related projects mainly focus on statistical analysis and visualization of pandemic data.

iii. Web applications. In the red part of Figure 4, this kind of project is mainly used to track the pandemic situation, providing various forms and real-time data and its visual web application (web app). Some projects implement relevant functions from worldwide data monitoring to the distribution of pandemic data in small communities. In addition to providing data tracking and presentation, some web application projects also include information collection and communication features. Examples include the collection of patient symptoms and a centralized communication platform for medical staff and volunteers.

iv. Mobile applications. In the yellow part of Figure 4, these projects mainly provide some mobile applications. Some of the mobile application projects have similar functionality to web applications in that they track and visualize real-time dynamic information about the pandemic (coronavirus tracking). The most popular assignments in this category implement proximity tracking and exposure notification. The primary function of these projects is to use the Bluetooth of mobile devices to record exposure information and thus notify the user of any contact with the patient.

In summary, the GitHub community of practice projects that started during the pandemic wanted to address the following three main areas: web and mobile app-based data information services, applications around all phases of the data science lifecycle, and the application of AI.

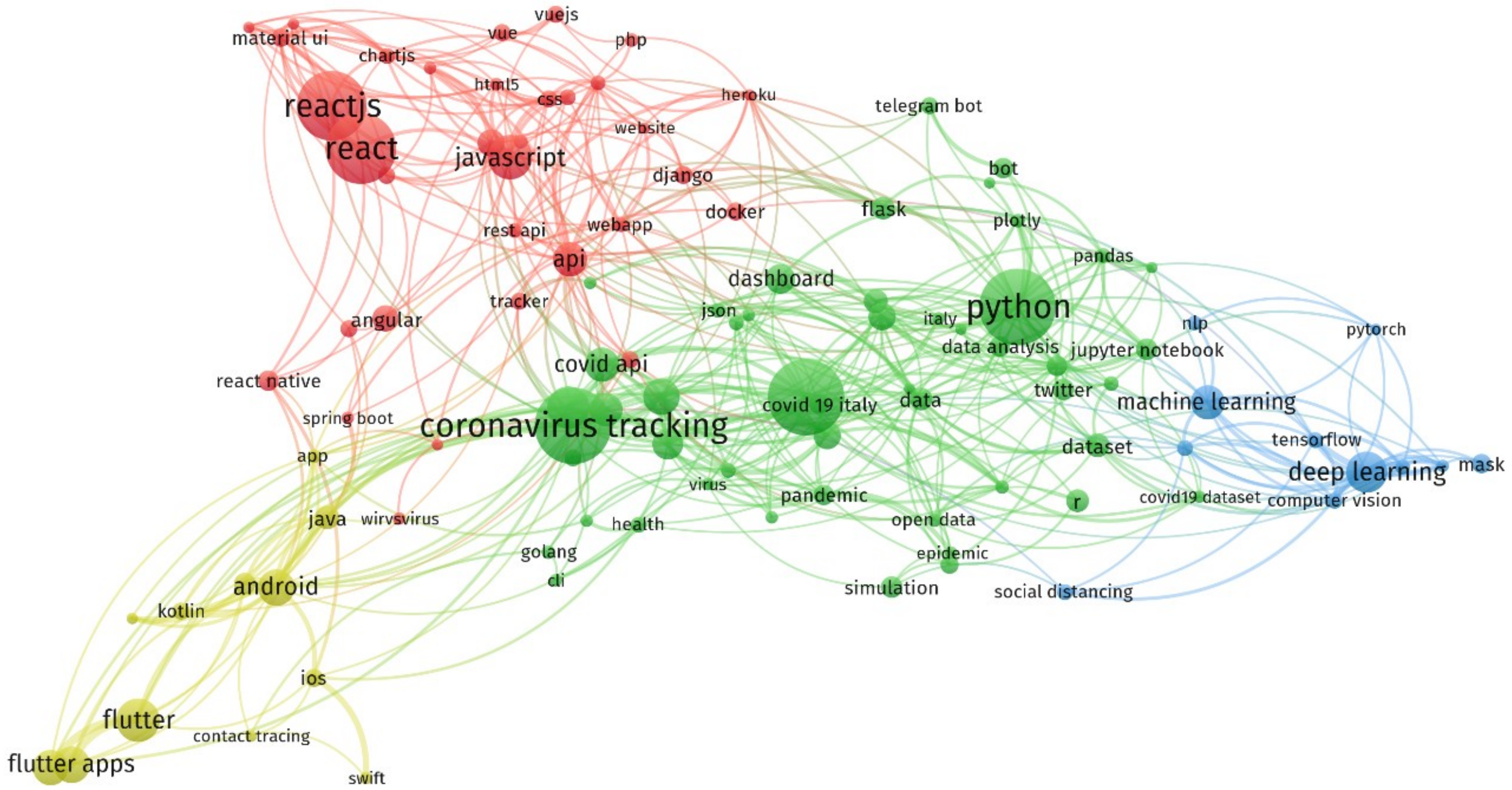

Figure 5. Clustering results of core hashtags of COVID-19 related projects.

### 4.3 Trends of COVID-19 related projects

The average word frequency share of hashtags in the above four categories of functionalities was counted, and the results are shown in Figure 6. As seen in Figure 6, the most significant number of COVID-19-related projects are apps. Apps include mobile apps and web apps, which account for more than 50%. Projects for data science and AI applications account for 30% and 18%.

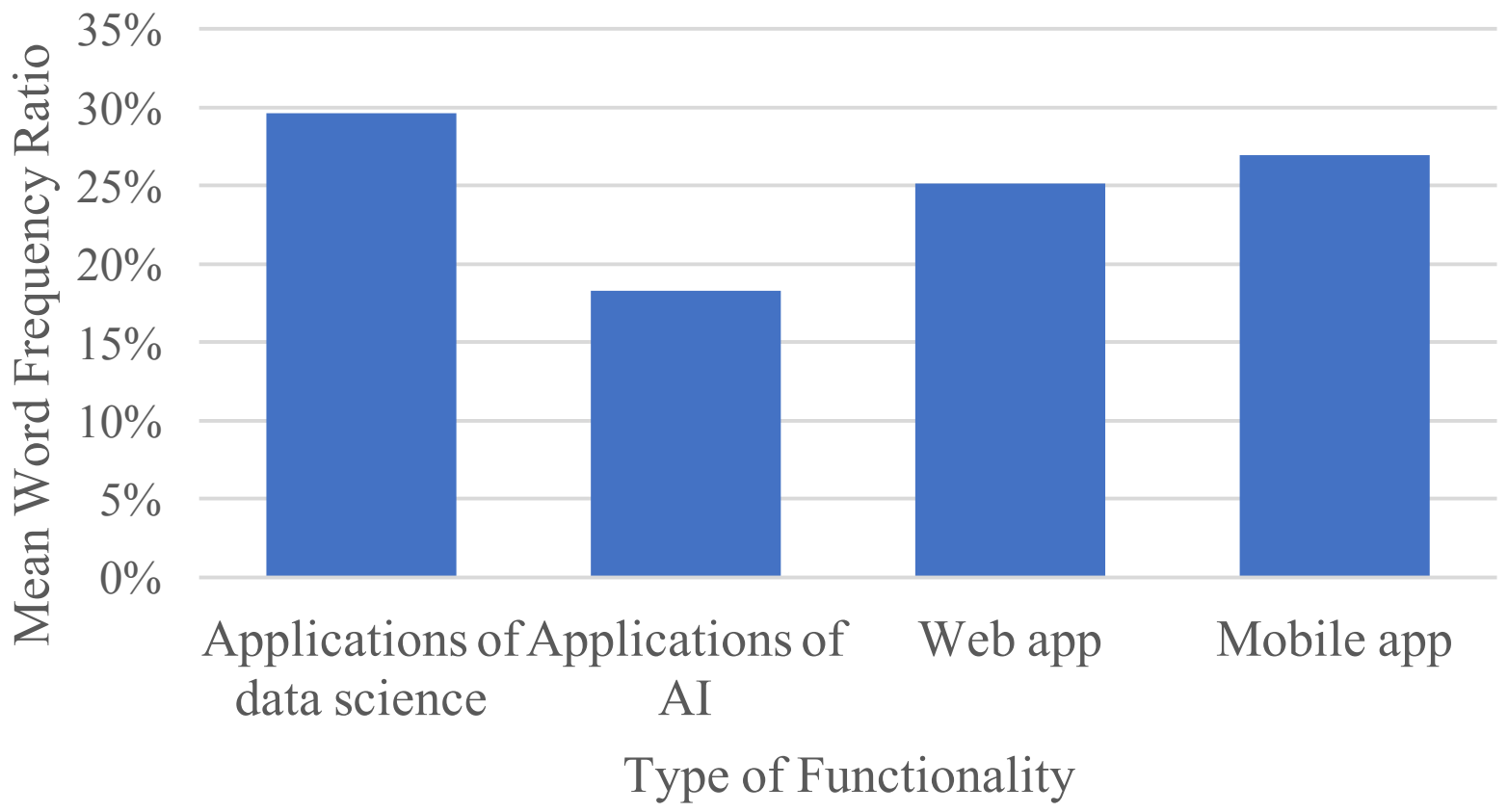

Figure 6. Average word frequency of hashtags for different functionalities.

The average word frequency of the hashtags in the four categories was counted by month, and the results are shown in Figure 7. An interesting conclusion can be drawn from the figure, i.e., the projects in the three different functionalities have essentially the same trend of change except for mobile applications. Mobile applications show another direction, probably related to contact tracking applications. In April, the focus on mobile applications peaked when Google and Apple announced reliable support for this type of program on their operating systems. The trend of change is the same for all kinds of projects and may be related to the skills possessed by developers within the platform. Most developers have only partial skills. For example, mobile application developers tend to focus only on mobile application development and have no other skills. The skill composition of developers within the community of practice is relatively stable. Since the number of developers with each skillset is regular, the trend of change in projects across functional types is also steady. Understandably, data science-related applications reached a high peak earlier. Both applications and AI applications need to be built on a data foundation. Only after the initial data collection and storage are completed can a range of applications such as data information services be performed.

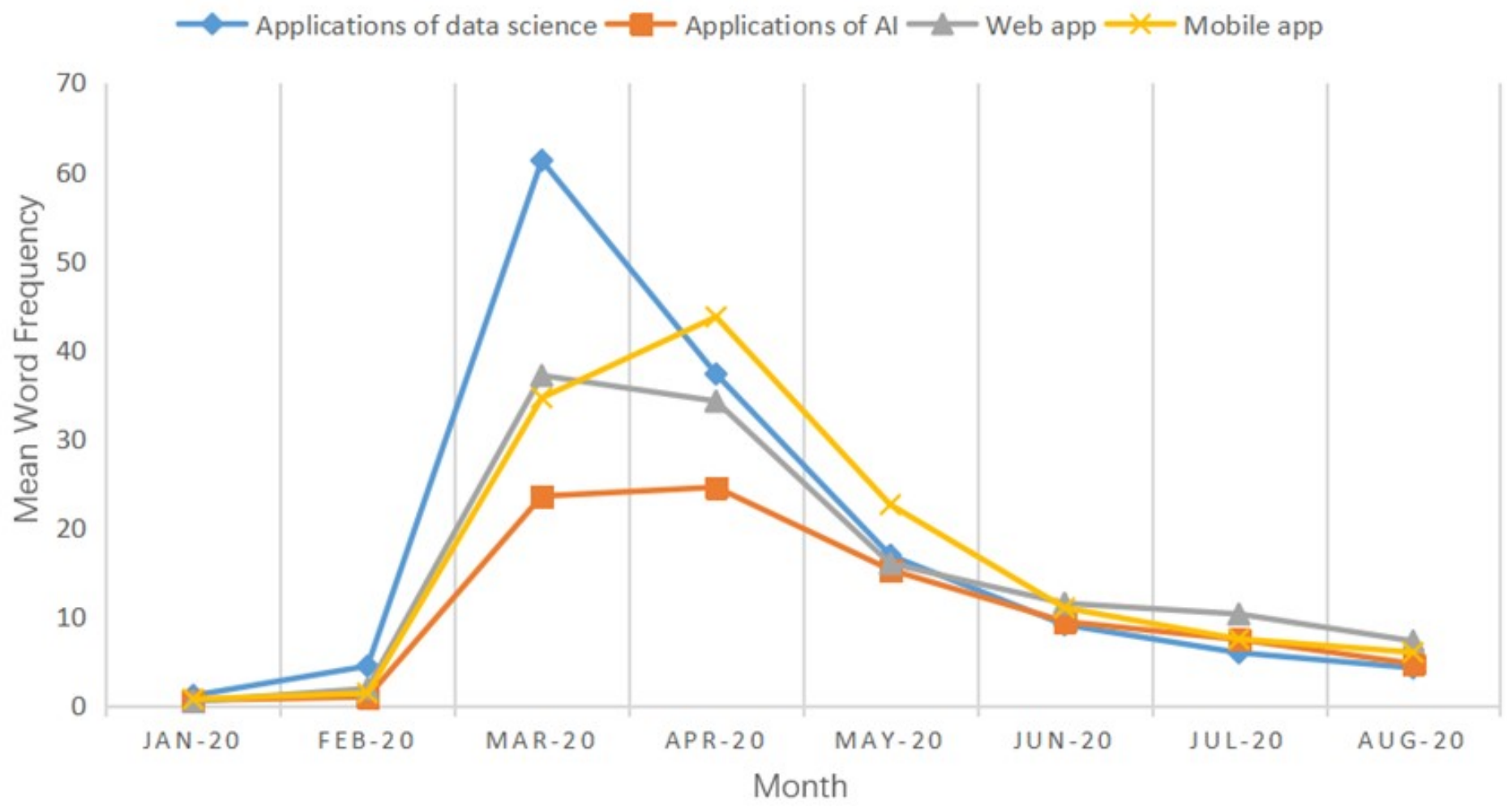

Figure 7. Trends in hashtag frequency evolution of projects for different functionalities.

### 4.4 Association rules between functionalities and technologies

The minimum support and minimum confidence were set to 0.002 and 0.2. To reduce the complexity, only the frequent binomial set was extracted. Data science and AI application projects are highly correlated, so association rules are removed for the hashtags of data science and AI projects. The network diagram is drawn based on the obtained association rules with a lift greater than 1.

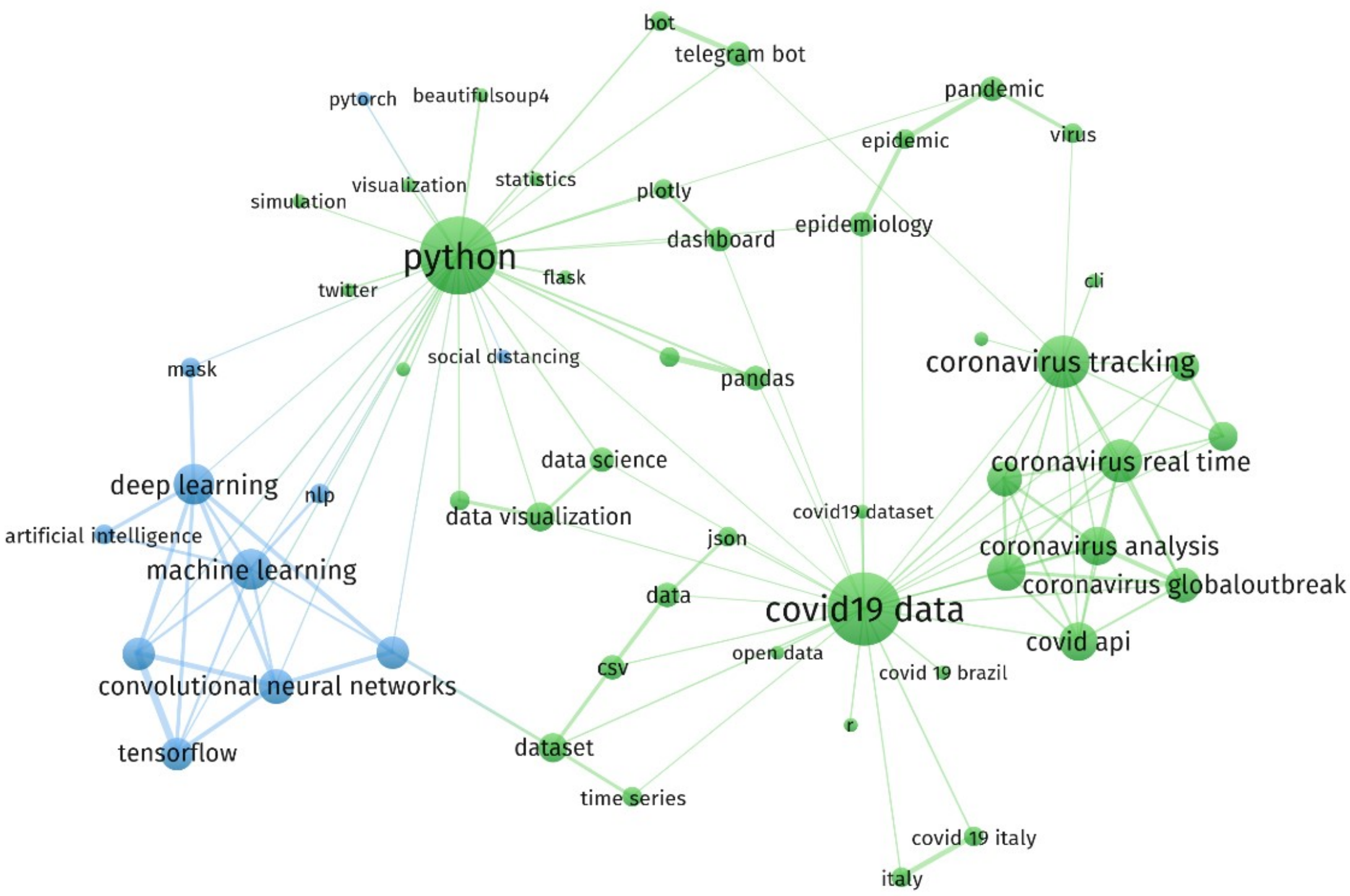

Figure 8. Association rules network of data science and AI application projects in GitHub.

Figure 8 shows the association rules between hashtags in data science and AI-related applications. The blue part in the figure is the hashtags for AI-related projects, and the green part is the hashtags for data science related projects. Each node in the graph represents a hashtag. The larger the node, the higher the word frequency. The connection between nodes means that there are association rules between them. The thickness of the link reflects the lift of the association rule. The thicker the link, the higher the lift degree. Whether it is a data science application or an AI-related application, Python is the core programming language that deserves to be used in such applications. The technologies associated with it include various python libraries and frameworks. The R language also has a place in this category, but its use is mainly limited to data analysis. As shown in Figure 8, the pandemic-related data (covid19 data) focuses on openness (open data), and the data storage formats are mainly CSV and JSON. At the same time, these data focus on the timeliness of epidemic tracking (Coronavirus real time). In applying artificial intelligence, deep learning and machine learning are the core technologies in AI-related projects. In applications related to computer vision, convolutional neural networks (CNN) are closely associated with it. TensorFlow is the most commonly used framework for AI-related projects on GitHub, and Pytorch is also used, but less so than TensorFlow. This is different from the academic community, where Pytorh has been used more often in the last two years[13]. This is a reflection of the differences between the academic and practice communities.

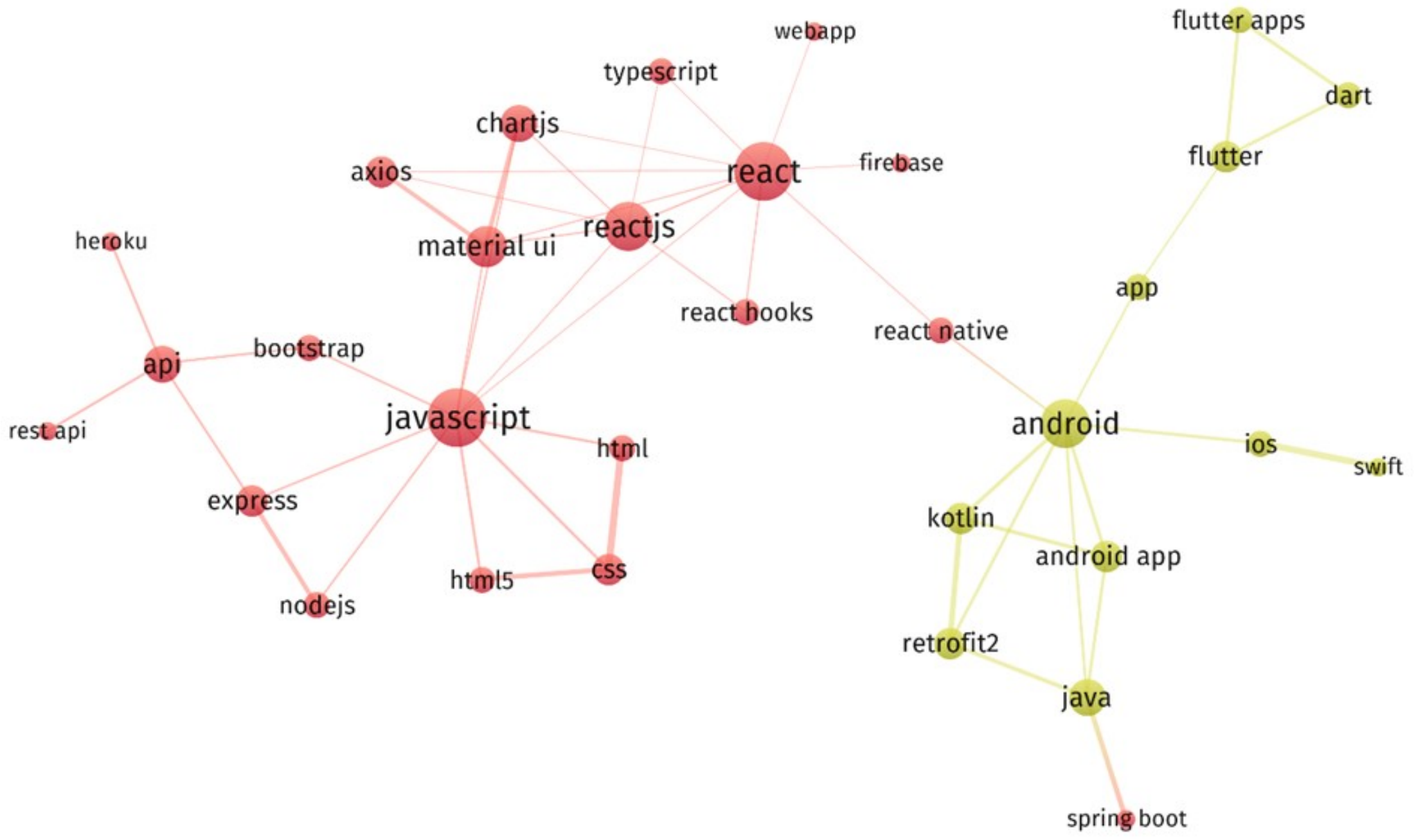

Figure 9. Application projects' hashtags association rules network in GitHub about COVID-19.

Figure 9 shows the association rules between hashtags in the application project, where red is the hashtag of the web application and yellow is the hashtag of the mobile application. This kind of project occupies an important position in the COVID-19 related projects of GitHub. From Figure 9, we can determine the mainstream programming languages and frameworks used by developers in the community of practice for Android and IOS development today. Android developers in the community of practice mostly use Java, dart, and Kotlin as programming languages. Swift is the only programming language highly related to IOS development. The technology related to web applications shown in Figure 9 is mainly front-end technology. As you can see, the react framework occupies an essential position in front-end technology, and some JavaScript libraries can also be found widely used on GitHub.

## 5.0 Discussion and conclusion

This study reviews the functionalities of COVID-19-related projects in the IT community of practice in the pandemic and the associations between functionalities and technologies. This section will discuss three aspects of IT applications during the pandemic, the association between functionalities and technologies and implications for practice.

### 5.1 The application of technology in communities of practice during the pandemic

Our findings suggest that projects submitted in the GitHub community during the pandemic were mainly related to data information services, outbreak trend prediction, and diagnostic treatment of the outbreak. The public needs to be informed of the current outbreak transmission status in the pandemic and access and communicate information. Infectious disease scientists need outbreak data to make predictions about the spread and development of an outbreak. Biomedical researchers need patient data to find ways to diagnose and treat patients. An essential foundation for achieving these needs is data, including outbreak statistics (statistical information on confirmed diagnoses, deaths, and cures), patient clinical data, etc. However, the performance of the emerging technology is not yet stable due to insufficient COVID-19 data sets, non-uniform data formats, missing data, and noise (Mbunge et al. 2021). As you can see, data is an important foundation for technology applications during the pandemic. Looking at the projects submitted on GitHub, there was a rapid increase in data science related projects at the beginning of the pandemic. This phenomenon reflects the quick response of the community to the pandemic.

The evolution of the different projects on GitHub shows that the various functional open-source projects have the same evolutionary trends. This may reflect the skill composition of the people in the community of practice. The submission of projects in the community is partly demand-driven of the pandemic but more likely related to the developers' skills in the community. At the time developers join the community, they possess specific skills. This makes them prefer or only implement certain applications. In a community of practice, these preferences and tendencies will influence community members (developers) to create and absorb new knowledge (Roberts, 2006). Communities of practice are groups of people who share expertise and are passionate about a common cause, connected informally (Wenger & Snyder, 2000). Developers develop and upload related projects spontaneously, and this spontaneous lack of organized development activity may not meet actual demand quickly enough.

### 5.2 Association between functionalities and technologies

This study explores the correlation between project functionalities and technologies through open-source project hashtags correlation analysis. It provides a macro perspective on the current state of adoption of new technologies in the GitHub community of practice. The GitHub community has the highest percentage of apps in its pandemic-related projects. The main programming languages used in developing these applications include Java, Dart, Flutter, JavaScript, etc. These technologies are widely accepted and used in the community of practice, and there are many developers in the community who have mastered the skill. These technologies have a high impact and a high level of technology maturity. Other vital projects in the community are related applications based on data science and artificial intelligence. The technologies involved in these projects mainly include data modeling, predictive analytics, machine learning, and deep learning. Artificial intelligence applications occupy the smallest share of the community. The application of artificial intelligence provides some innovative solutions to combat the pandemic, and this type of technology is more cutting-edge and an important direction for innovation.

The survey of the community of practice obtained different results than the survey of the academic literature. A review of the use of modern technology in the pandemic through the academic literature reveals more discussion of cutting-edge technologies. These technologies mainly include artificial intelligence (AI), telemedicine, blockchain, 5G, Internet of Things (IoT), etc. (e.g., Alharbi & Abdur Rahman, 2021; Mbunge et al., 2021; Vaishya et al., 2021). And a survey of the community of practice revealed that more projects launched during the pandemic were applications. These programs are more likely to provide data information services, and the technology used is more mature.

## 5.3 Implications for practice

All the findings show that emerging technologies are widely used in the pandemic. The pandemic has received widespread attention from developers in the IT community of practice. Data is the cornerstone of the rapid application of emerging technologies. In the early stages of a pandemic, the first thing that relevant developers and scientists need to accomplish should be collecting, storing, and sharing various types of pandemic data. In addition, in the big data environment, we should consider transforming open data into open link data and analyzing the information, conducive to the rapid development and use of applications (Victorino et al. 2018). During the pandemic, the application of technology in the community of practice differs significantly from the application in the academic community. Researchers and developers from different disciplines should collaborate more, which will help knowledge transfer and facilitate the application of emerging technology (Teixeira et al., 2019).

The current GitHub community of developers submitting open-source projects is more likely to be a spontaneous act. Communities of practice should provide relevant services in future crises and organize community members to contribute their skills efficiently. Improving the quality of open-source projects in a pandemic requires organizational management by the community. In the late stages of the pandemic, communities of practice such as GitHub should organize pandemic-related projects more effectively to provide a large pool of pre-existing technologies for responding to possible future crises. For example, use hierarchical tags to manage related projects. The hierarchical topic structure can provide users with more comprehensive and precise topics (Zhang et al. 2019).

## 6.0 Limitations and future work

The research in this paper has some shortcomings. On the one hand, only relevant projects on GitHub are explored and analyzed. To understand the status of open-source projects during the pandemic, more open source communities may need to be mined and analyzed. On the other hand, it is difficult to understand the true intent of developers through the econometric analysis of projects submitted by developers. Future work may require questionnaires or interview methods to collect data to get a clearer picture of developer intent. In addition, future work could analyze maps of the community from multiple perspectives and provide a more fine-grained analysis of the use of technology. Also, we can consider the study of the collaboration between project development teams and explore the collaboration perspective between developers and scientists.

## Note

1. https://docs.github.com/en/free-pro-team@latest/github/getting-started-with-github/github-glossary#readme
2. https://docs.github.com/en/free-pro-team@latest/github/administering-a-repository/classifying-your-repository-with-topics#about-topics
3. https://developer.github.com/v3/search/
4. There are 61 languages in the data collected in this paper, of which English is the most, accounting for 84%. In addition, Portuguese, Spanish, and Chinese account for 3.3%, 2.7%, and 2%, respectively.
5. http://api.fanyi.baidu.com
6. https://www.ranks.nl/stopwords
7. https://radimrehurek.com/gensim/
8. https://radimrehurek.com/gensim/models/word2vec.html
9. https://scikitlearn.org/stable/modules/generated/sklearn.feature_extraction.text.TfidfVectorizer.html#sklearn.feature_extraction.text.TfidfVectorizer
10. https://scikitlearn.org/stable/modules/generated/sklearn.cluster.SpectralClustering.html#sklearn.cluster.SpectralClustering
11. https://www.vosviewer.com/
12. None of the projects in the table has a hashtag, and the survey date is December 4, 2020
13. https://paperswithcode.com/trends

## Acknowledgement

This work is supported by the National Natural Science Foundation of China (No. 72074113, No. 71974095).

## References

Adams, Eric C, and Christopher Freeman. 2000. "Communities of Practice: Bridging Technology and Knowledge Assessment." *Journal of Knowledge Management* 4: 38-44. https://doi.org/10.1108/13673270010315939.

Agrawal, Rakesh, and Ramakrishnan Srikant. 1994. "Fast Algorithms for Mining Association Rules in Large Databases." *In Proceedings of the 20th International Conference on Very Large Data Bases 12-15 September 1994 Santiago, Chile,* edited by Jorge B. Bocca, Matthias Jarke, and Carlo Zaniolo, 487–99. VLDB '94. San Francisco, CA, USA: Morgan Kaufmann Publishers Inc.

Alharbi, Ayman, and Md Abdur Rahman. 2021. "Review of Recent Technologies for Tackling COVID-19." *SN Computer Science* 2: 460. https://doi.org/10.1007/s42979-021-00841-z.

Alshrari, Ahmed S., Shuaibu A. Hudu, Mohd Imran, Syed Mohammed Basheeruddin Asdaq, Alreshidi M. Ali, and Syed Imam Rabbani. 2022. "Innovations and Development of Covid-19 Vaccines: A Patent Review." *Journal of Infection and Public Health* 15: 123–31. https://doi.org/10.1016/j.jiph.2021.10.021.

Apostolopoulos, Ioannis D., and Tzani A. Mpesiana. 2020. "Covid-19: Automatic Detection from X-Ray Images Utilizing Transfer Learning with Convolutional Neural Networks." *Physical and Engineering Sciences in Medicine* 43: 635–40. https://doi.org/10.1007/s13246-020-00865-4.

Barge-Gil, Andræs. 2013. "Open Strategies and Innovation Performance." *Industry & Innovation*

20: 585–610. https://doi.org/10.1080/13662716.2013.849455.

Brem, Alexander, Eric Viardot, and Petra A. Nylund. 2021. "Implications of the Coronavirus (COVID-19) Outbreak for Innovation: Which Technologies Will Improve Our Lives?" *Technological Forecasting and Social Change* 163: 120451. https://doi.org/10.1016/j.techfore.2020.120451.

Breiman, Leo. 2001. "Random Forests." *Machine Learning* 45: 5–32. https://doi.org/10.1023/A:1010933404324.

Calinski, T., and J. Harabasz. 1974. "A Dendrite Method for Cluster Analysis." *Communications in Statistics - Theory and Methods* 3: 1–27. https://doi.org/10.1080/03610927408827101.

Chakraborty, Amrita, and Arpan Kumar Kar. 2021. "How Did COVID-19 Impact Working Professionals–a Typology of Impacts Focused on Education Sector." *The International Journal of Information and Learning Technology 3*8: 273-282. https://doi.org/10.1108/IJILT-06-2020-0125.

Chang, Chih-Chung, and Chih-Jen Lin. 2011. "LIBSVM: A Library for Support Vector Machines." *ACM Transactions on Intelligent Systems and Technology* 2: 1–27. https://doi.org/10.1145/1961189.1961199.

Chen, Ji-De, and Hung-Yu Kao. 2015. "LDA Based Semi-Supervised Learning from Streaming Short Text." *In 2015 IEEE International Conference on Data Science and Advanced Analytics (DSAA) 19-21 October 2015 Paris, France*. USA: IEEE, 1-8. https://doi.org/10.1109/DSAA.2015.7344830.

Ghosh, Indranil, and Manas K Sanyal. 2021. "Introspecting Predictability of Market Fear in Indian Context during COVID-19 Pandemic: An Integrated Approach of Applied Predictive

Modelling and Explainable AI." *International Journal of Information Management Data Insights* 1 (2): 100039. https://doi.org/10.1016/j.jjimei.2021.100039.

He, Wu, Zuopeng (Justin) Zhang, and Wenzhuo Li. 2021. "Information Technology Solutions, Challenges, and Suggestions for Tackling the COVID-19 Pandemic." *International Journal of Information Management* 57: 102287. https://doi.org/10.1016/j.ijinfomgt.2020.102287.

Hosmer, David W., Stanley Lemeshow, and Rodney X. Sturdivant. 2013. *Applied Logistic Regression. 3rd ed.* Wiley Series in Probability and Statistics. Wiley. https://doi.org/10.1002/9781118548387.

Jakhar, Deepak, and Ishmeet Kaur. 2020. "Current Applications of Artificial Intelligence for COVID - 19." *Dermatologic Therapy* 33. https://doi.org/10.1111/dth.13654.

Javaid, Mohd, Abid Haleem, Raju Vaishya, Shashi Bahl, Rajiv Suman, and Abhishek Vaish. 2020. "Industry 4.0 Technologies and Their Applications in Fighting COVID-19 Pandemic." *Diabetes & Metabolic Syndrome: Clinical Research & Reviews* 14: 419–22. https://doi.org/10.1016/j.dsx.2020.04.032.

Jianbo Shi, and J. Malik. 2000. "Normalized Cuts and Image Segmentation." *IEEE Transactions on Pattern Analysis and Machine Intelligence* 22: 888–905. https://doi.org/10.1109/34.868688.

Karami, Amir, Brandon Bookstaver, Melissa Nolan, and Parisa Bozorgi. 2021. "Investigating Diseases and Chemicals in COVID-19 Literature with Text Mining." *International Journal of Information Management Data Insights* 1: 100016. https://doi.org/10.1016/j.jjimei.2021.100016.

Keestra, Sarai, Florence Rodgers, Rhiannon Osborne, and Sabrina Wimmer. 2022. "University Patenting and Licensing Practices in the United Kingdom during the First Year of the

COVID-19 Pandemic.” *Global Public Health* 17: 641–51. https://doi.org/10.1080/17441692.2022.2049842.

Keller, James M., Michael R. Gray, and James A. Givens. 1985. “A Fuzzy K-Nearest Neighbor Algorithm.” *IEEE Transactions on Systems, Man, and Cybernetics* SMC-15: 580–85. https://doi.org/10.1109/TSMC.1985.6313426.

Koch, Julian, Ralf Plattfaut, and Ingo Kregel. 2021. “Looking for Talent in Times of Crisis–The Impact of the Covid-19 Pandemic on Public Sector Job Openings.” *International Journal of Information Management Data Insights* 1: 100014. https://doi.org/10.1016/j.jjimei.2021.100014.

Manning, Christopher D., Prabhakar Raghavan, and Hinrich Schütze. 2008. *Introduction to Information Retrieval*. New York: Cambridge University Press.

Marsick, Victoria J, Andrew K Shiotani, and Martha A Gephart. 2014. “Teams, Communities of Practice, and Knowledge Networks as Locations for Learning Professional Practice.” *In International Handbook of Research in Professional and Practice-Based Learning*, 1021–41. Springer.

Marzal, A., and E. Vidal. 1993. “Computation of Normalized Edit Distance and Applications.” *IEEE Transactions on Pattern Analysis and Machine Intelligence* 15: 926–32. https://doi.org/10.1109/34.232078.

Mbunge, Elliot, Boluwaji Akinnuwesi, Stephen G. Fashoto, Andile S. Metfula, and Petros Mashwama. 2021. “A Critical Review of Emerging Technologies for Tackling COVID - 19 Pandemic.” *Human Behavior and Emerging Technologies* 3: 25–39. https://doi.org/10.1002/hbe2.237.

Mikolov, Tomas, Kai Chen, Greg Corrado, and Jeffrey Dean. 2013. "Efficient Estimation of Word Representations in Vector Space." ArXiv:1301.3781 [Cs], September. http://arxiv.org/abs/1301.3781.

Pickerill, Peter. 2020. "PHANTOM: Curating GitHub for Engineered Software Projects Using Time-Series Clustering." *Empirical Software Engineering*, 25: 2897-2929. https://doi.org/10.1007/s10664-020-09825-8.

Roberts, J. (2006). Limits to communities of practice. *Journal of Management Studies*, 43, 623–639.

Randhawa, Gurjit S., Maximillian P. M. Soltysiak, Hadi El Roz, Camila P. E. de Souza, Kathleen A. Hill, and Lila Kari. 2020. "Machine Learning Using Intrinsic Genomic Signatures for Rapid Classification of Novel Pathogens: COVID-19 Case Study." Edited by Oliver Schildgen. *PLOS ONE* 15: e0232391. https://doi.org/10.1371/journal.pone.0232391.

Srinivasa Rao, Arni S. R., and Jose A. Vazquez. 2020. "Identification of COVID-19 Can Be Quicker through Artificial Intelligence Framework Using a Mobile Phone–Based Survey When Cities and Towns Are under Quarantine." *Infection Control & Hospital Epidemiology* 41: 826–30. https://doi.org/10.1017/ice.2020.61.

Salton, Gerard, and Christopher Buckley. 1988. "Term-Weighting Approaches in Automatic Text Retrieval." *Information Processing & Management* 24: 513–23. https://doi.org/10.1016/0306-4573(88)90021-0.

Shrestha, Prasha, Arun Sathanur, Suraj Maharjan, Emily Saldanha, Dustin Arendt, and Svitlana Volkova. 2020. "Multiple Social Platforms Reveal Actionable Signals for Software Vulnerability Awareness: A Study of GitHub, Twitter and Reddit." Edited by Sergi Lozano. PLOS ONE 15 (March): e0230250. https://doi.org/10.1371/journal.pone.0230250.

Soos, Carlin, and Gregory H. Leazer. 2020. "Presentations of Authorship in Knowledge Organization." *KNOWLEDGE ORGANIZATION* 47: 486–500. https://doi.org/10.5771/0943-7444-2020-6-486.

Solanki, Vikas, Umesh Solanki, Anupum Baliyan, Vinay Kukreja, Vikas Lamba, and Bidush Kumar Sahoo. 2021. "Importance of Artificial Intelligence and Machine Learning in Fighting with COVID-19 Epidemic." *In 2021 5th International Conference on Information Systems and Computer Networks (ISCON), Mathura, India*: IEEE, 1–8. https://doi.org/10.1109/ISCON52037.2021.9702316.

Song, Ligang, and Yixiao Zhou. 2020. "The COVID-19 Pandemic and Its Impact on the Global Economy: What Does It Take to Turn Crisis into Opportunity?" *China & World Economy* 28: 1–25.

Teixeira, S. J., Veiga, P. M., & Fernandes, C. A. (2019). The knowledge transfer and cooperation between universities and enterprises. *Knowledge Management Research & Practice*, 17, 449–460. https://doi.org/10.1080/14778238.2018.1561166

Victorino, Marcio, Maristela Terto de Holanda, Edison Ishikawa, Edgard Costa Oliveira, and Sammohan Chhetri. 2018. "Transforming Open Data to Linked Open Data Using Ontologies for Information Organization in Big Data Environments of the Brazilian Government: The Brazilian Database Government Open Linked Data – DBgoldbr." *KNOWLEDGE ORGANIZATION* 45: 443–66. https://doi.org/10.5771/0943-7444-2018-6-443.

Vaishya, Raju, Mohd Javaid, Ibrahim Haleem Khan, Abhishek Vaish, and Karthikeyan P Iyengar. 2021. "Significant Role of Modern Technologies for COVID-19 Pandemic." *Journal of*

*Industrial Integration and Management* 06: 147–59. https://doi.org/10.1142/S242486222150010X.

Wang, Qiang, Min Su, Min Zhang, and Rongrong Li. 2021. "Integrating Digital Technologies and Public Health to Fight Covid-19 Pandemic: Key Technologies, Applications, Challenges and Outlook of Digital Healthcare." *International Journal of Environmental Research and Public Health* 18: 6053. https://doi.org/10.3390/ijerph18116053.

Wenger, E. C., & Snyder, W. M. (2000). Communities of practice: The organizational frontier. *Harvard Business Review*, 78, 139–146. https://doi.org/10.1108/13673270010315939。

Wit-de Vries, Esther de, Wilfred A. Dolfsma, Henny J. van der Windt, and M. P. Gerkema. 2019. "Knowledge Transfer in University–Industry Research Partnerships: A Review." *The Journal of Technology Transfer* 44: 1236–55. https://doi.org/10.1007/s10961-018-9660-x.

Yadav, Hitesha, Arpan Kumar Kar, and Smita Kashiramka. 2021. "How Does Entrepreneurial Orientation and SDG Orientation of CEOs Evolve before and during a Pandemic." *Journal of Enterprise Information Management* 35: 160-178. https://doi.org/10.1108/JEIM-03-2021-0149.

Zhang, Chengzhi, Hua Zhao, Xuehua Chi, and Shuitian Ma. 2019. "Information Organization Patterns from Online Users in a Social Network." *KNOWLEDGE ORGANIZATION* 46: 90–103. https://doi.org/10.5771/0943-7444-2019-2-90.

Zhang, Min-Ling, and Zhi-Hua Zhou. 2014. "A Review on Multi-Label Learning Algorithms." *IEEE Transactions on Knowledge and Data Engineering* 26: 1819–37. https://doi.org/10.1109/TKDE.2013.39.

Zhao, Zhuo, Yangmyung Ma, Adeel Mushtaq, Abdul M. Azam Rajper, Mahmoud Shehab, Annabel

Heybourne, Wenzhan Song, Hongliang Ren, and Zion Tsz Ho Tse. 2021. “Applications of Robotics, Artificial Intelligence, and Digital Technologies During COVID-19: A Review.” *Disaster Medicine and Public Health Preparedness*, 1–11. https://doi.org/10.1017/dmp.2021.9.